# Accelerated infection testing at scale: a proposal for inference with single test on multiple patients

**March 30, 2020**

**Tarun Jain**, Indian Institute of Management Ahmedabad, Email: tj9d@virginia.edu

**Bijendra Nath Jain**, Indraprastha Institute of Information Technology Delhi and Indian Institute of Technology Delhi, Email: bnjain@iiitd.ac.in

## 1. Introduction

In pandemics or epidemics, public health authorities need to rapidly test a large number of individuals, both to determine the line of treatment as well as to know the spread of infection to plan containment, mitigation and future responses. However, the lack of adequate testing kits could be a bottleneck, especially in the case of unanticipated new diseases, such as COVID-19, where the testing technology, manufacturing capability, distribution, human skills and laboratories might be unavailable or in short supply. In addition, the cost of the standard PCR test is approximately USD 48, which is prohibitive for poorer patients and most governments. We address this bottleneck by proposing a test methodology that pools the sample from two (or more) patients in a single test. The key insight is that a single negative result from a pooled sample likely implies negative infection of all the individual patients. and It thereby rules out further tests for the patients. This protocol, therefore, requires significantly fewer tests. This may, however, result in somewhat increased false negatives. Our simulations show that combining samples from two patients with 7% underlying likelihood of infection implies that 36% fewer test kits are required, with 14% additional units of time for testing.

## 2. Proposal

We illustrate our proposal for the case of testing for COVID-19, which uses nasal swabs to collect samples, which are then placed in a slot in an RNA extractor. The turnaround time to obtain results from the standard PCR-based test for COVID-19 is anywhere from 24 hours to several days (Daley 2020), while the actual time taken by RNA extractor may be in the order of an hour or two. We shall term the latter time duration as "1 unit" of time.

**Protocol 1.** The standard testing procedure uses a single kit for a single patient. Hence, the straight forward approach to determine whether individual members of a group of K=2 persons, {P1, P2}, are infected is as follows.

1. Test P1 and publish the outcome, and

2. Test P2 and publish the outcome.

Clearly, this protocol requires exactly two test kits and uses two slots in an RNA extractor that can extract RNA from multiple swabs. In that case the test duration (= time spent by swabs in the lab) is 1 unit. For K persons, the number of kits required is K. It uses K slots in the RNA extractor, and the resulting test duration is 1 unit (assuming K is small enough).

**Protocol 2.** The proposed protocol pools portions from two swabs to create one sample, with the balance from each swab saved separately.

1. Use one kit to test whether the pooled sample (taken earlier from the two swabs) is infected.
2. If the result is negative, the conclusion is that both individuals, P1 and P2, have tested negative for the infection. This outcome, viz. P1 and P2 are both negative, can be published.
3. Else if the test is positive:
   - Test P1 for infection using the (balance of) P1's swab, and publish the result, and
   - Test P2 for infection using the (balance of) of P2's swab, and publish the result.

Similar protocols have been previously applied, for example malaria testing (Zhou et al 2014) and even in designing protocols for computer networks (Tanenbaum & Wetherall 2011).[1]

## 3. Simulations

**Main calculations**

First, we report results from a simulation with simple parameters: K=2, the underlying independent probability of infection is 2%, 7% and 12%. For the present, we assume that test accuracy is 100%, with no false positives or false negatives. Each test takes one unit of time.

Subsequent simulations will allow for a broader range of parameters.

In Table 1, consider the case where the infection probability is 7%.

$$\Pr(P1 \text{ is negative}) = 0.93, \Pr(P2 \text{ is negative}) = 0.93$$

The pooled sample will show a negative result only if both are negative.

$$\Pr(P1 \text{ and } P2 \text{ are negative}) = 0.93*0.93 = 0.8649$$

---

[1] The suggested protocol has its basis in "random-walk protocol" that attempts to identify the number of wireless stations ready to transfer frames and determine the order in which the ready stations may send frames.

If the test is positive, which will happen in the remaining 13.51% cases, then P1 and P2 should both be independently tested.

The average number of kits required is (1 kit * 0.8649 cases) + (3 kits * 0.1351 cases) = 1.27 kits, which saves 36% kits compared to the standard protocol.

**Table 1. Two-person example**

|  | Number of test kits | Probability person is infected | | | | | |
|---|---|---|---|---|---|---|---|
|  |  | 2% | | 7% | | 12% | |
| Event (2 persons) |  | Probability of event | No. of kits used | Probability of event | No. of kits used | Probability of event | No. of kits used |
| P1 and P2 negative | 1 | 96% | 0.96 | 86.49% | 0.86 | 77.44% | 0.77 |
| P1 or P2 or both are positive | 3 | 4% | 0.12 | 13.51% | 0.41 | 22.56% | 0.68 |
| Average kits required |  |  | 1.08 |  | 1.27 |  | 1.45 |
| Savings, % of kits |  |  | (46%) |  | (36%) |  | (27%) |
| Average total test time (units) |  |  | 1.04 |  | 1.14 |  | 1.23 |
| Average increase in time (units) |  |  | 4% |  | 14% |  | 23% |

The savings in the number of kits is dramatically greater – 46% - with a 2% infection rate. This rate is *comparable* to the COVID-19 infection rate reported in Wuhan, China (50,800 cases out of a population of 11 million, or 4.6 cases per thousand). Correspondingly, the savings in the number of kits is lower as infection rates increase, since a larger fraction of cases will require three kits instead of one. Even with 12% of the population infected, 27% fewer kits are required under our proposed protocol.

The trade-off is in the time taken to deliver results to physicians and/or patients. The standard protocol takes one unit of time (of the RNA extractor, and of lab personnel) if both tests are run simultaneously. The proposed protocol is a two-step process where the second set of samples are tested only if the first test in step 1 is positive. In the above example (with infection rate of 7%), the second test is run in 14% of cases, which is also the increase in the test time. The increase in average time is lower for 2% infection rates – only 4% increase in time for which lab resources are used. The total time taken to obtain a swab sample, transport it to lab for testing, and return results to patient or physician may not be impacted very much.

The probability of the second round test increases as the infection probabilities increase (see Figure 1). This decreases test kit savings and increases the additional time required for tests. Figure 1 shows that with these parameters, the test kit savings is positive for infection rates lower than 30%. No credible estimate for COVID-19 predicts such high simultaneous infection rates – the highest infection rate for the Spanish Influenza was approximately one-in-three, but spread over a number of years.

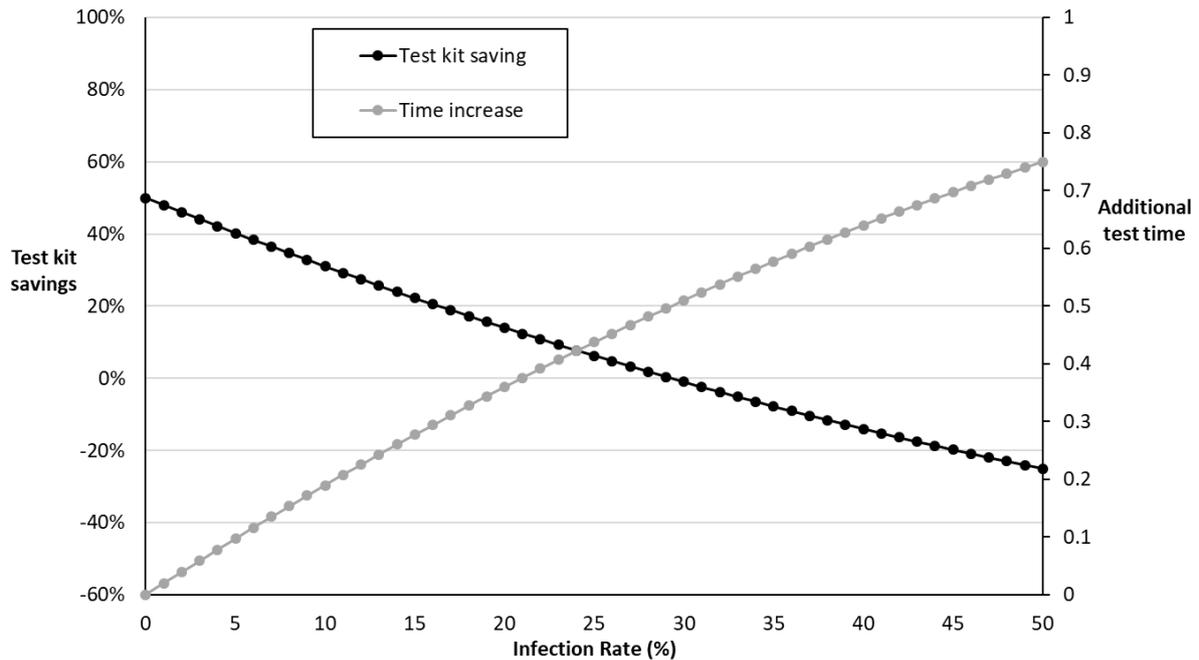

**Figure 1: Test kit saving and Additional test time**

**Efficiency enhancement**

One method to increase efficiency of Protocol 2 is to pool more than two samples in a test. For K = 3, the modified protocol pools the sample from three patients and a negative result is interpreted as negative for all three patients. Correspondingly, three tests are required in the second stage if the first stage is positive. Even so, a simulation with three patients and a 7% infection rate shows that 47% fewer test kits are required compared to 36% fewer kits for in the two patient case (K=2).

Another method for increasing efficiency is to pool samples from related patients – for instance, for highly contagious diseases such as COVID-19, a co-residing couple is likely to have the same infection status. Thus, a positive result can be interpreted as a positive result for both, alleviating the need for the second test, and further reducing the number of test kits.

**Test error**

We next incorporate test imperfections (false negatives and false positives) into our analysis. A person falsely tested negative incorrectly reassures her that she is infection free. Consequently, and unwittingly, she is more likely to spread the disease. In addition, public health authorities are less likely to trace contacts of individuals they believe are uninfected. This is particularly harmful in the early stages of the epidemic or pandemic when containing infections is the primary objective of public health.

False positives create a situation where a person is unnecessarily isolated and treated. This is harmful both to the patient who is undergoing treatment for the disease she does not have, and the rest of the public health system ends up misallocating scarce resources (such as hospital beds, physician and nurse time, drugs and other medical equipment) to treat individuals who did not need treatment and deprive others that need treatment.

To understand the influence of test errors, both false negatives and false positives, we carried out simulations of Protocol 1 and Protocol2. We simulated 50,000 patient pairs, assuming the following parameters:

- The true infection rate is 7% (or 10%),
- The rate of false positives returned by the test is 3% (or 5%), and
- The rate of false negatives returned by the test is 5% (or 10%, 15% and 20%).

In other words, a total of 16 simulations were run for Protocol 1 and for Protocol 2, as shown in Table 2 below.

**Table 2: Simulations for evaluating false negatives and false positives**

|  | Infection rate is 7% | | Infection rate is 10% | |
| --- | --- | --- | --- | --- |
|  | False positive rate is 3% | False positive rate is 5% | False positive rate is 3% | False positive rate is 5% |
| False negative rate is 5% | Simulation with 50,000 patient pairs | Simulation with 50,000 patient pairs | Simulation with 50,000 patient pairs | Simulation with 50,000 patient pairs |
| False negative rate is 10% | Simulation with 50,000 patient pairs | Simulation with 50,000 patient pairs | Simulation with 50,000 patient pairs | Simulation with 50,000 patient pairs |
| False negative rate is 15% | Simulation with 50,000 patient pairs | Simulation with 50,000 patient pairs | Simulation with 50,000 patient pairs | Simulation with 50,000 patient pairs |
| False negative rate is 20% | Simulation with 50,000 patient pairs | Simulation with 50,000 patient pairs | Simulation with 50,000 patient pairs | Simulation with 50,000 patient pairs |

We first describe the simulation run for Protocol 1: If a patient P1 is truly infected with probability 0.07, she is declared negative *falsely* with probability 5%, else (correctly) declared positive. If P1 is not infected, with probability 0.93, she is declared positive *falsely* with probability 3%, else (correctly) declared negative. And similarly for patient P2.

In the simulation for Protocol 2, if persons P1 and P2 are both truly negative, then the simulation returns (in step 1) a positive test result *falsely* with probability 3%. Else, the simulation (correctly) declares P1 and P2 to be uninfected. In the former case, P1 and P2 are tested separately (and the simulation method of Protocol 1 applies). In the latter case, no further testing is required.

In case P1 or P2, or both, are truly positive, the simulation returns in step 1 a negative test result *falsely* with probability 5%, and no further testing is required. Else, when the test returns in step 1 a positive result, P1 and P2 are tested separately and the simulation method of Protocol 1 applies.

In both cases the results are tabulated by counting the number of times person P1 is correctly (or falsely) declared infected or uninfected to obtain the following confusion matrix.

**Table 3: Confusion matrix**

| | Protocol 1 simulation<br>Positive rate 7%<br>False positive rate 3%<br>False negative rate 5% | | | Protocol 2 simulation<br>Positive rate = 7%<br>False positive rate = 3%<br>False negative rate = 5% | |
|---|---|---|---|---|---|
| | Tested Positive | Tested Negative | | Tested Positive | Tested Negative |
| Truly Positive | 3403 | **171** | Truly Positive | 3229 | **345** |
| Truly Negative | **1400** | 45026 | Truly Negative | **134** | 46292 |

Table 3 shows that false negatives increase from 171 in Protocol 1 (5% rate) to 345 in Protocol 2 (9.77% rate). False positives decrease from 1400 (3% rate) to 134 (0.28%). Tables 4A and Table 4B show the fraction of false negative results from following Protocol 2, and Tables 4C and Table 4D show the fraction of false positive results from following Protocol 2.

**Table 4A: False negative rate with 7% infection rate**

Infection rate 7%

**Resulting false negative rate from Protocol 2**

| | False positive rate | |
|---|---|---|
| False negative rate | *3%* | *5%* |
| *5%* | 10% | 10% |
| *10%* | 19% | 19% |
| *15%* | 28% | 29% |
| *20%* | 36% | 36% |

**Table 4B: False negative rate with 10% infection rate**

Probability of infection 10%

**Resulting false negative rate from Protocol 2**

| | False positive rate | |
|---|---|---|
| False negative rate | *3%* | *5%* |
| *5%* | 10% | 9% |
| *10%* | 19% | 19% |
| *15%* | 28% | 28% |
| *20%* | 36% | 36% |

**Table 4C: False positive rate with 7% infection rate**

Probability of infection 7%

**Resulting false positive rate from Protocol 2**

| False negative rate | False positive rate | |
|---|---|---|
| | *3%* | *5%* |
| *5%* | 0.28% | 0.56% |
| *10%* | 0.28% | 0.55% |
| *15%* | 0.26% | 0.53% |
| *20%* | 0.25% | 0.51% |

**Table 4D: False positive rate with 10% infection rate**

Probability of infection 10%

**Resulting false positive rate from Protocol 2**

| False negative rate | False positive rate | |
|---|---|---|
| | *3%* | *5%* |
| *5%* | 0.37% | 0.71% |
| *10%* | 0.35% | 0.67% |
| *15%* | 0.32% | 0.65% |
| *20%* | 0.32% | 0.62% |

## 4. Conclusion

We propose a testing protocol to accelerate infection diagnostics by combining multiple samples, and retesting individual samples only in the case of positive results. The key insight is that a negative result in the first stage implies negative infection for all individuals, which means that a single test could rule out infection in multiple individuals. We show that this protocol reduces the required number of testing kits, which alleviates a key bottleneck for public health authorities in times of pandemics and epidemics such as COVID-19. Our proposed protocol is relatively more effective when the infection rate is low, which suggests that it is better suited for early stage, as well as large scale, population-wide testing. We summarize the trade off in terms of additional time for testing, which might be an important operational or diagnostic consideration.

**Acknowledgement**

The authors wish to acknowledge the contribution of Tushit Jain of San Diego, CA in creating the simulator to assess impact of Protocol 2 on false negatives/positives, and for his help in interpreting the results.